\documentclass[journal,twocolumn]{IEEEtran}

\usepackage{amsmath,amsfonts,amssymb,amsthm}
\usepackage{cite}
\usepackage{graphicx}
\usepackage{subfig}
\usepackage{caption}

\usepackage{xcolor}

\usepackage{epstopdf}		
\usepackage{algorithm}
\usepackage{algpseudocode}
\usepackage{balance}

\newcommand{\tr}{\mathrm{Tr}} 
\newcommand{\V}{\mathrm{Var}} 
\newcommand{\myT}{{\mathsf{T}}} 
\newcommand{\myH}{{\mathsf{H}}} 
\newcommand{\diag}{{\rm{diag}}} 

\title{Phase-Aware Localization in Pinching Antenna Systems: CRLB Analysis and ML Estimation}
\author{Hao Feng, Ebrahim Bedeer, Ming Zeng, Xingwang Li, Shimin Gong, and Quoc-Viet Pham 
    \thanks{H. Feng is with Hunan Institute of Engineering, and Donghua University as well as Laval University (email: 1219001@mail.dhu.edu.cn).}

    \thanks{E. Bedeer is with University of Saskatchewan, Saskatoon, SK, Canada (email: e.bedeer@usask.ca).}
    
    \thanks{M. Zeng is with Laval University, Quebec City, Canada (email: ming.zeng@gel.ulaval.ca).}

    \thanks{X. Li is with Henan Polytechnic University, Jiaozuo, China (email: lixingwang@hpu.edu.cn).}

    \thanks{S. Gong is with Sun Yat-sen University, Guangzhou, China (e-mail: gongshm5@mail.sysu.edu.cn).}

    \thanks{Q.-V. Pham is School of Computer Science and Statistics, Trinity College Dublin, Dublin 2, D02PN40, Ireland (e-mail: Viet.Pham@tcd.ie).}

}

\begin{document}
\maketitle

\begin{abstract}
Pinching antenna systems (PASS) have emerged as a promising architecture for high-frequency wireless communications. In this letter, we investigate user localization in PASS by jointly exploiting the received signal amplitude and phase information. A complex baseband signal model is formulated to capture free-space path loss, waveguide attenuation, and distance-dependent phase rotation between the user and each pinching antenna. Based on this model, we derive the Fisher information matrix and closed-form Cramer--Rao lower bound and position error bound. {\color{black}The derived analysis reveals that the phase-induced Fisher information decays with the fourth power of the user--antenna distance, whereas the amplitude-induced information decays with the sixth power, explaining the fundamental advantage of phase-aware localization in typical PASS deployments.} A maximum likelihood estimator is then developed and {\color{black}implemented} through a two-stage procedure combining coarse grid search and Levenberg--Marquardt refinement. Numerical results show that the proposed estimator achieves low positioning error and generally outperforms the considered benchmarks under different noise powers, numbers of pinching antennas, and user locations.
In the considered scenario, the proposed method achieves sub-meter-level accuracy over the evaluated service area and yields substantially lower positioning error than the amplitude-only benchmark.
\end{abstract}


\section{Introduction}\label{sec:Introduction}
Pinching antenna systems (PASS) have emerged as a promising architecture for high-frequency wireless communications, particularly in millimeter-wave and terahertz bands \cite{Atsushi_22,  liu2025pinching}. By enabling reconfigurable antenna ports along a dielectric waveguide, PASS provides spatial flexibility for line-of-sight (LoS) restoration, spectral/energy-efficiency enhancement, and adaptive beam control \cite{ding2024, Zeng_COMML25, Zhao_TCOM25, zeng2025EE, Tegos_2025}. Most existing PASS studies assume perfect user location and channel state information (CSI), which are difficult to obtain in practice. Since PASS channels are strongly geometry-dependent due to free-space loss, waveguide attenuation, and distance-sensitive phase rotation, inaccurate user localization can directly degrade communication and resource-allocation performance.

Recent efforts have relaxed these idealized assumptions. Channel estimation techniques for PASS have been investigated in \cite{Xiao_COMML25, Zhou_SPAWC25}, and robust resource allocation schemes under user location uncertainty have been proposed in \cite{zeng2025robust, feng2026, Sun_TVT26}. \textcolor{black}{Meanwhile, PASS has recently attracted increasing attention for localization and sensing applications. In \cite{zhang2025}, user positioning in PASS was studied based on received signal strength (RSS). In \cite{He_WCL26}, PASS-assisted localization was investigated from a stochastic-geometry perspective, where RSS measurements were used to characterize localizability and the Cramér–Rao lower bound (CRLB) distribution. In parallel, PASS-assisted integrated sensing and communication (ISAC) was studied in \cite{Kumar_TWC26}, where the communication outage probability and sensing mean-square error were analyzed.} \textcolor{black}{However, existing PASS localization studies mainly rely on RSS or amplitude-related measurements, while the phase information contained in the complex baseband signal has not been fully exploited for direct coordinate estimation.} In LoS-based systems, phase variations are highly sensitive to propagation distance, suggesting that amplitude-only approaches are intrinsically suboptimal from an estimation-theoretic standpoint.

Motivated by these observations, this work develops a comprehensive framework for user localization in PASS that jointly exploits both received signal amplitude and phase. {\color{black}Although phase-aided localization is well established, its role in PASS is distinct because the received complex signal is jointly determined by free-space propagation, waveguide attenuation, and pinching antenna (PA) positions. Hence, both the amplitude and phase carry PASS-specific geometric and waveguide-dependent information that is not captured by amplitude-only localization models.}
The main contributions of this paper are summarized as follows:
\begin{itemize}
   \item {\bf{Unified Signal Model for Localization in PASS}:} We establish a rigorous signal model that incorporates free-space path loss, waveguide attenuation, and phase rotation. This model explicitly reveals how user position affects both amplitude and phase of the received signal across sequentially activated pinching antennas.
    \item {\bf{Closed-form Characterization and Design Insights}:} We derive the Fisher information matrix (FIM) and obtain closed-form expressions for the CRLB and the position error bound (PEB). {\color{black}By decomposing the per-PA Fisher information into amplitude- and phase-induced components, we show that the phase-induced information decays more slowly with the user--antenna distance than the amplitude-induced one.
   This explains why phase-aware localization is particularly beneficial in PASS and provides design insights into the effects of PA placement, user geometry, and waveguide attenuation.}
    \item {\bf{Two-Stage Maximum Likelihood (ML) Localization Algorithm:}} We develop a phase-aware ML estimator and {\color{black}address} the resulting non-convex problem via coarse grid search followed by Levenberg--Marquardt (LM) refinement. Numerical results show substantial accuracy gains over existing benchmarks.
\end{itemize}


\section{System Model}\label{sec:system_model}
We consider an uplink communication scenario in which $K$ single-antenna users communicate with an access point (AP) through $N$ PAs deployed on a single dielectric waveguide. The $K$ users are uniformly distributed  on the ground plane within a rectangular area of dimension $\ell_1 \times \ell_2$ meters. The system operates using time division multiplexing, where only one user transmits in a given time slot. The AP sequentially activates the $N$ PAs to receive the signal from the $k$th user to help with the localization as explained later in this section. 
The sequential PA scan is assumed fast enough such that the user location remains fixed and the noise samples are independent.

We assume the feed point is located at ${\mathbf{v}}_0 = [0, 0, d]^\myT$, and the dielectric waveguide, of length $\ell_{\rm{w}} \leq \ell_2$, is located along the $y$-axis at height $d$ above the ground plane. Hence, the location of the $n$th PA, $n = 1, ..., N,$ is denoted as ${\mathbf{v}}_n = [0, v_n, d]^\myT$ where $v_n \in [0, \ell_{\rm{w}}]$.  The $K$ users are located on the ground floor, i.e., $z = 0$. Hence, the 2D location of the $k$th user, $k = 1, ..., K,$ is $\mathbf{u}_k = [u_{x,k}, u_{y,k}]^\myT$. We assume that the user location  $\mathbf{u}_k$ is unknown and needs to be estimated.

The combined channel $h_{k,n}$ from  user $k$ to the AP consists of 1) the large-scale channel, 2) the small-scale fading, and 3) the attenuation inside the dielectric waveguide. The large-scale channel is mainly due to free-space loss and under LoS assumption it is given by \cite{ding2024}
\begin{IEEEeqnarray}{RCL}
	h^{(\rm{ls})}_{k,n}(d_{k,n}) & = & \frac{\lambda}{4 \pi d_{k,n}},
\end{IEEEeqnarray}
where $\lambda$ is the wavelength and $d_{k,n}$ is the distance from the $k$th user to the $n$th PA, which is given by
\begin{IEEEeqnarray}{RCL}
	d_{k,n} & = & \sqrt{u_{x,k}^2 + (u_{y,k} - v_n)^2 + d^2}. 
\end{IEEEeqnarray} 
The small-scale fading is modeled as
\begin{IEEEeqnarray}{RCL}
	h^{(\rm{ss})}_{k,n}(d_{k,n}) & = & \gamma_{k,n} e^{-{j 2 \pi d_{k,n}}/{\lambda}},
\end{IEEEeqnarray}
where $\gamma_{k,n}$ is the channel gain. 
The attenuation inside the dielectric waveguide is modeled as \cite{rao2015microwave}
\begin{IEEEeqnarray}{RCL}
	h^{(\rm{w})}_{n} & = & e^{-(\alpha + j \beta) v_n},
\end{IEEEeqnarray}
where $\alpha = \pi \sqrt{\epsilon_r} \tan(\delta)/ \lambda$ and $\beta = 2 \pi \sqrt{\epsilon_r}/\lambda$, with $\epsilon_r$ and $\tan(\delta)$ representing the relative permittivity and loss angle tangent, respectively \cite{rao2015microwave}.

The received signal at the AP coming from the $k$th user through the $n$th PA is given by
\begin{IEEEeqnarray}{RCL}
	r_{k,n} & = & h^{(\rm{ls})}_{k,n}(d_{k,n}) h^{(\rm{ss})}_{k,n}(d_{k,n}) h^{(\rm{w})}_{n} \sqrt{p_k} s_k + w_{k,n} \nonumber \\
		& = & \frac{\gamma_{k,n} a_{k,n}}{d_{k,n}} e^{-{j 2 \pi d_{k,n}}/{\lambda}} s_k + w_{k,n}, \quad \forall k,
\end{IEEEeqnarray}
where $a_{k,n} = \lambda \sqrt{p_k}  e^{-(\alpha + j \beta) v_n}/ 4 \pi$, $p_k$ and $s_k$ are the transmit power and the unit-power transmit pilot symbol, respectively, and $w_{k,n}$ is the zero-mean additive white Gaussian noise (AWGN) with variance $\sigma^2$. Following \cite{zhang2025}, we assume that the channel gain is normalized, i.e., $\gamma_{k,n} = 1$. {\color{black}
The present work focuses on a LoS-dominant and ideally synchronized setting to establish a fundamental phase-aware localization benchmark for PASS; robust localization under multipath propagation, residual synchronization errors, channel uncertainty, phase noise, and hardware impairments is left for future work.}

Given that the noise samples $w_{k,n}$, $n = 1, ..., N$, are independent and identically distributed, the joint probability density function (PDF) of all received signals from all $N$ PAs for the user $k$ can be expressed as
\begin{IEEEeqnarray}{RCL}
	f_{\mathbf{{u}_k}}(\mathbf{r}_k|\mathbf{{u}_k}) &=& \frac{1}{(\pi \sigma^2)^{N}} \exp\left(-\frac{\Vert\mathbf{r}_k - \mathbf{s}_k(\mathbf{{u}_k})\Vert^2}{\sigma^2} \right),
\end{IEEEeqnarray}
where $\mathbf{r}_k = [r_{k,1}, ..., r_{k,N}]^\myT$ and $\mathbf{s}_k(\mathbf{{u}_k}) = [({\gamma_{k,1} a_{k,1}}/{d_{k,1}}) e^{-{j 2 \pi d_{k,1}/{\lambda}}} s_k, ..., ({\gamma_{k,N} a_{k,N}}/{d_{k,N}}) e^{-{j 2 \pi d_{k,N}/{\lambda}}} s_k]^\myT$. Hence, the log-likelihood function $L_{\mathbf{r}_k}(\mathbf{u}_k)$ is given as
\begin{IEEEeqnarray}{RCL}\label{eq:likelihood}
	L_{\mathbf{r}_k}(\mathbf{u}_k) & \propto & -\frac{1}{\sigma^2} (\mathbf{r}_k - \mathbf{s}_k(\mathbf{{u}_k}))^\myH(\mathbf{r}_k - \mathbf{s}_k(\mathbf{{u}_k})),
\end{IEEEeqnarray}
where $(\cdot)^\myH$ denotes the conjugate transpose operation. 

{\color{black}
The proposed method adopts direct coordinate-based localization, where the Cartesian user location $\mathbf u_k=[u_{x,k},u_{y,k}]^\myT$ is directly estimated from the complex received signal vector $\mathbf r_k$. Unlike two-step methods based on intermediate parameters such as angle/time of arrival, or range, it directly fits the PASS-specific signal model $\mathbf s_k(\mathbf u_k)$ to the received observations.
}

\section{CRLB Derivation and Analysis}\label{sec:CRLB}
The FIM can be written as
\begin{IEEEeqnarray}{RCL}
	\mathbf{J}(\mathbf{u}_k) &=& \frac{2}{\sigma^2} \Re\left\{ 
	\left( \frac{\partial \mathbf{s}_k(\mathbf{u}_k)}{\partial \mathbf{u}_k} \right)^\mathsf{H} 
	\frac{\partial \mathbf{s}_k(\mathbf{u}_k)}{\partial \mathbf{u}_k} 
	\right\},
\end{IEEEeqnarray}
where $\frac{\partial \mathbf{s}_k(\mathbf{u}_k)}{\partial \mathbf{u}_k} \in \mathbb{C}^{N \times 2}$ is the Jacobian matrix of the signal vector $\mathbf{s}_k(\mathbf{u}_k)$ with respect to the $k$th user location $\mathbf{u}_k$. Using the chain rule, we obtain
\begin{IEEEeqnarray}{RCL}\label{eq:partial_1}
	\frac{\partial s_{k,n}}{\partial u_{x,k}} 
		 = & a_{k,n} s_k e^{-j\frac{2\pi}{\lambda} d_{k,n}}\left(-\frac{u_{x,k}}{d_{k,n}^3}-j\frac{2\pi}{\lambda}\frac{u_{x,k}}{d_{k,n}^2}\right),
\end{IEEEeqnarray}
Similarly, one can calculate
\begin{IEEEeqnarray}{RCL}\label{eq:partial_2}
	\frac{\partial s_{k,n}}{\partial u_{y,k}} 
		& = & a_{k,n} s_k e^{-j\frac{2\pi}{\lambda} d_{k,n}}\left(-\frac{u_{y,k}-v_n}{d_{k,n}^3}-j\frac{2\pi}{\lambda}\frac{u_{y,k}-v_n}{d_{k,n}^2}
\right). \nonumber \\
\end{IEEEeqnarray}

Equations \eqref{eq:partial_1} and \eqref{eq:partial_2} can be rewritten, respectively, as
\begin{IEEEeqnarray}{RCL}
	\frac{\partial s_{k,n}}{\partial u_{x,k}} & = & - m_{k,n}(d_{k,n}) u_{x,k}, \\
	\frac{\partial s_{k,n}}{\partial u_{y,k}} & = & - m_{k,n}(d_{k,n}) (u_{y,k}-v_n),
\end{IEEEeqnarray}
where $m_{k,n}(d_{k,n}) = a_{k,n} s_k e^{-j\frac{2\pi}{\lambda} d_{k,n}}\left(\frac{1}{d_{k,n}^3}+j\frac{2\pi}{\lambda}\frac{1}{d_{k,n}^2}\right)$. Hence,
\begin{IEEEeqnarray}{RCL}
	\frac{\partial s_{k,n}}{\partial \mathbf{u}_k} & = & - m_{k,n}(d_{k,n}) \left[u_{x,k}, \quad (u_{y,k}-v_n)\right],
\end{IEEEeqnarray}
and 
\begin{IEEEeqnarray}{RCL}
	\frac{\partial \mathbf{s}_k(\mathbf{u}_k)}{\partial \mathbf{u}_k} & = & - \mathbf{M}_k(d_{k,n}) \: \mathbf{G}(\mathbf{u}_k), \label{eq:jac}
\end{IEEEeqnarray}
where $\mathbf{G}(\mathbf{u}_k) = [\mathbf{g}_1^\myT; ...; \mathbf{g}_N^\myT] \in \mathbb{R}^{N \times 2}$, $\mathbf{g}_n^\myT = [u_{x,k}, (u_{y,k}-v_n)]$, and $\mathbf{M}_k(d_{k,n}) = \diag\left(m_{k,1}(d_{k,1}), ..., m_{k,N}(d_{k,N})\right) \in \mathbb{C}^{N \times N}$, with $\diag(\cdot)$ denoting the diagonal matrix. Here, $\mathbf{M}(d_{k,n})$ captures the amplitude and phase sensitivities with respect to distance $d_{k,n}$, while $\mathbf{G}(\mathbf{u}_k)$ captures the geometric sensitivity of the distance with respect to the user location $\mathbf{u}_k$. Hence, the FIM can be reformulated as
\begin{IEEEeqnarray}{RCL}
	\mathbf{J}(\mathbf{u}_k) &=& \frac{2}{\sigma^2}\Re\left\{\mathbf{G}(\mathbf{u}_k)^\myH  \: \tilde{\mathbf{M}}_k(d_{k,n}) \: \mathbf{G}(\mathbf{u}_k)  \right\},
\end{IEEEeqnarray}
where $\tilde{\mathbf M}_k(\mathbf d_{k,n})=\mathbf M_k^H(\mathbf d_{k,n})\mathbf M_k(\mathbf d_{k,n})$ is diagonal and contains the magnitude-squared distance sensitivities.

{\color{black}
To further interpret the FIM, define the per-PA information weight as
\begin{IEEEeqnarray}{RCL}
    \omega_{k,n} &=& |m_{k,n}(d_{k,n})|^2.
\end{IEEEeqnarray}
Since $\mathbf g_n^\myT=[u_{x,k},u_{y,k}-v_n]$, the FIM can be equivalently written as
\begin{IEEEeqnarray}{RCL}
    \mathbf J(\mathbf u_k)
    &=&
    \frac{2}{\sigma^2}
    \sum_{n=1}^{N}
    \omega_{k,n}\mathbf g_n\mathbf g_n^\myT .
    \label{eq:FIM_weighted_sum}
\end{IEEEeqnarray}
This expression shows that each PA contributes a rank-one information matrix, where $\omega_{k,n}$ measures the signal sensitivity and $\mathbf g_n$ captures the geometric sensitivity.
}

{\color{black}
Using the definition of $m_{k,n}(d_{k,n})$ and assuming $|s_k|=1$, the per-PA information weight is obtained as
\begin{IEEEeqnarray}{RCL}
    \omega_{k,n}
    &=&
    \frac{\lambda^2 p_k}{16\pi^2}e^{-2\alpha v_n}
    \left(
    \frac{1}{d_{k,n}^{6}}
    +
    \frac{4\pi^2}{\lambda^2}\frac{1}{d_{k,n}^{4}}
    \right).
    \label{eq:omega_decomposition}
\end{IEEEeqnarray}
The first term in \eqref{eq:omega_decomposition} comes from the derivative of the amplitude envelope $1/d_{k,n}$, whereas the second term originates from the derivative of the propagation phase $e^{-j2\pi d_{k,n}/\lambda}$. Thus,
\begin{IEEEeqnarray}{RCL}
    \omega_{k,n}
    =
    \omega_{k,n}^{(\mathrm{amp})}
    +
    \omega_{k,n}^{(\mathrm{ph})},
\end{IEEEeqnarray}
where
\begin{IEEEeqnarray}{RCL}
    \omega_{k,n}^{(\mathrm{amp})}
    &=&
    \frac{\lambda^2 p_k}{16\pi^2}e^{-2\alpha v_n}
    \frac{1}{d_{k,n}^{6}},
    \\
    \omega_{k,n}^{(\mathrm{ph})}
    &=&
    \frac{p_k}{4}e^{-2\alpha v_n}
    \frac{1}{d_{k,n}^{4}}.
\end{IEEEeqnarray}
Hence, the phase-induced information decays as $d_{k,n}^{-4}$, while the amplitude-induced one decays as $d_{k,n}^{-6}$. Their ratio is
\begin{IEEEeqnarray}{RCL}
    \frac{\omega_{k,n}^{(\mathrm{ph})}}
    {\omega_{k,n}^{(\mathrm{amp})}}
    =
    \frac{4\pi^2 d_{k,n}^2}{\lambda^2}.
    \label{ratio}
\end{IEEEeqnarray} 
For typical PASS deployments where $d_{k,n}\gg \lambda$, the phase-induced component dominates the Fisher information. This explains why exploiting the complex signal phase can substantially improve localization accuracy compared with amplitude-only PASS localization.
}

\begin{figure*}[htbp]
\centering
\subfloat[\label{fig:1a}]{\includegraphics[width=0.3\textwidth]{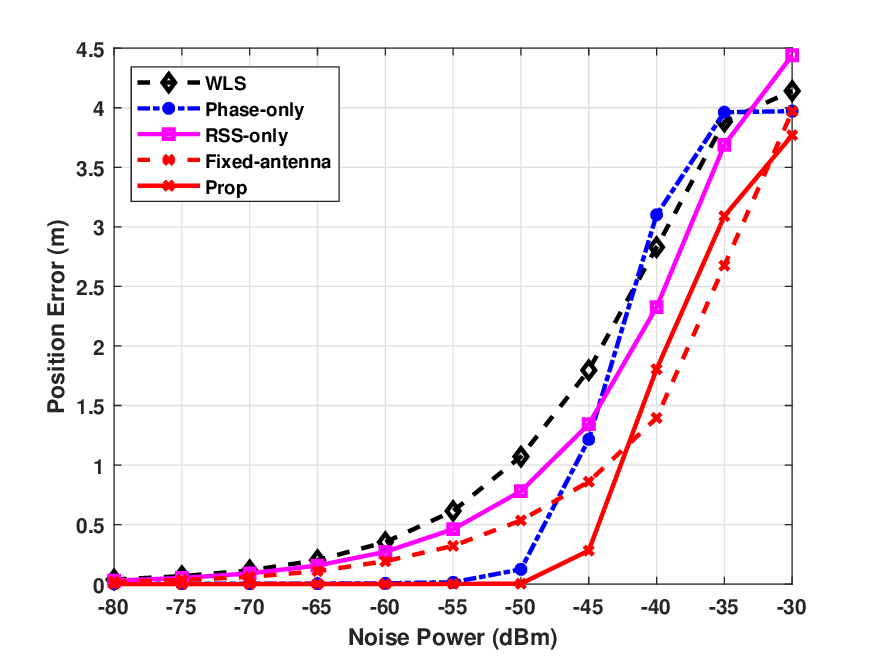}}\hfill
\subfloat[\label{fig:1b}] {\includegraphics[width=0.3\textwidth]{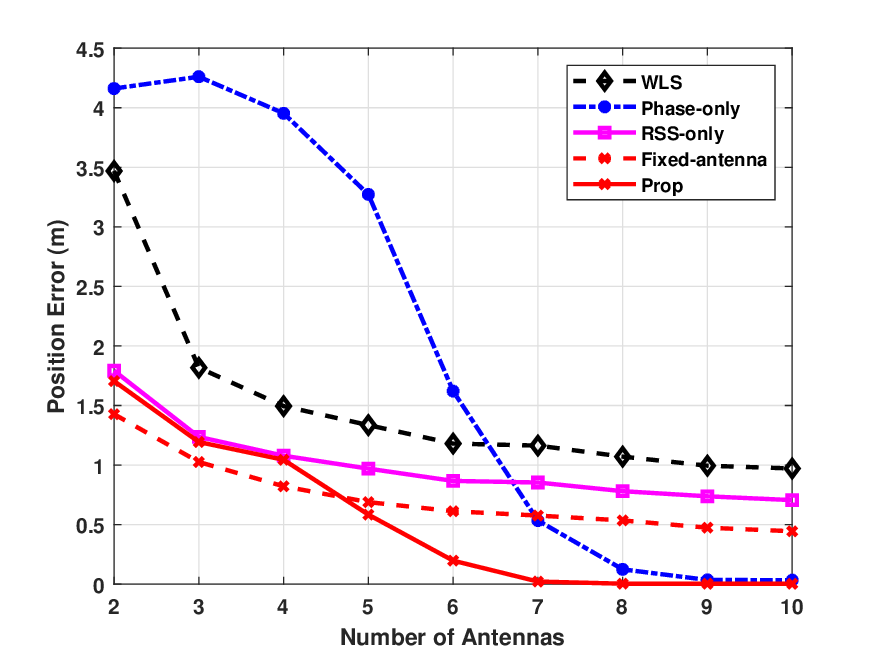}}\hfill
\subfloat[\label{fig:1c}]{\includegraphics[width=0.3\textwidth]{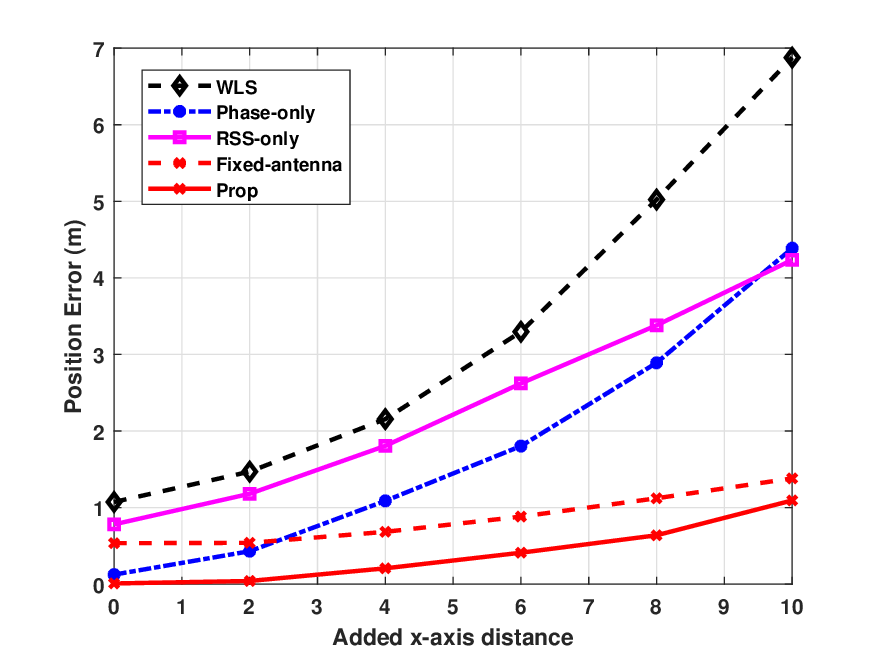}}
 \caption {{\color{black}Average position error (over 1000 random trials) versus: a) noise power; b) number of PAs; and c) added x-axis distance.}} \label{fig:performance}
\end{figure*}

\begin{figure}[t]
    \centering
    \includegraphics[width=0.7\linewidth]{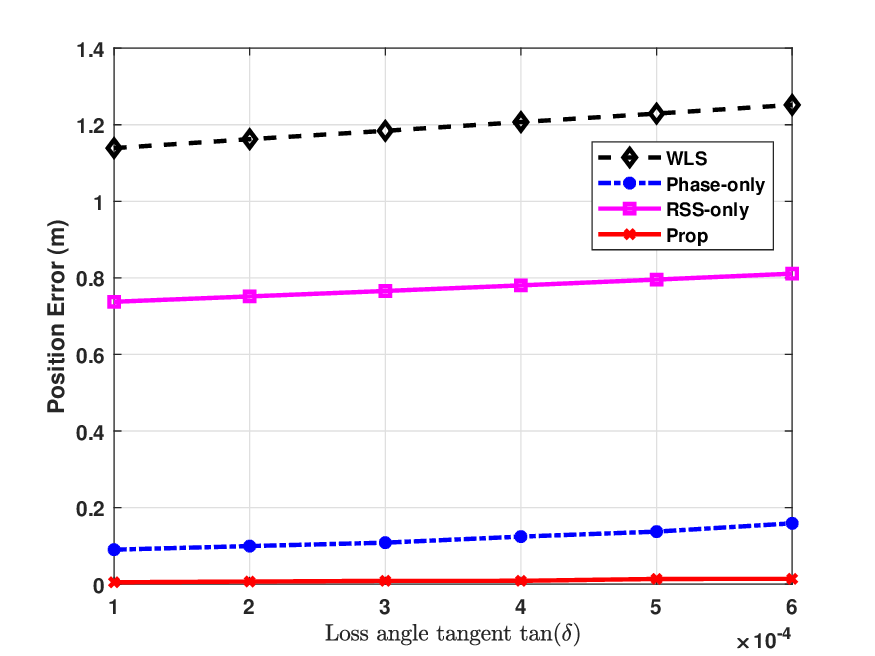}
    \caption{{\color{black}Average position error (over 1000 random trials) versus the waveguide loss angle tangent.}}
    \label{fig:waveguid loss}
\end{figure}

The CRLB for the unbiased estimator $\hat{\mathbf{u}}_k$ is given as
\begin{IEEEeqnarray}{RCL}
	{\rm{\mathbf{Cov}}}(\hat{\mathbf{u}}_k) 
			&\succeq& \frac{1}{{J}_{11} {J}_{22} - {J}^2_{12}} \begin{bmatrix}
				{J}_{22} & -{J}_{12}\\[4pt]
				-{J}_{21} & {J}_{11}
			\end{bmatrix},
\end{IEEEeqnarray}
where ${J}_{11}$, ${J}_{12}$, ${J}_{21}$, and ${J}_{22}$  are the block matrices of the FIM $\mathbf{J}(\mathbf{u}_k)$ whose elements are given as
\begin{IEEEeqnarray}{RCL}
	{J}_{11} & = & \frac{2}{\sigma^2}\sum_{n=1}^N |m_{k,n}(d_{k,n})|^2  u_{x,k}^2,  \\
	{J}_{12} = {J}_{21} & = & \frac{2}{\sigma^2}\sum_{n=1}^N |m_{k,n}(d_{k,n})|^2 u_{x,k}(u_{y,k}-v_n), \\
	{J}_{22} & = & \frac{2}{\sigma^2}\sum_{n=1}^N |m_{k,n}(d_{k,n})|^2 (u_{y,k}-v_n)^2.
\end{IEEEeqnarray}
{\color{black}
The above equations reveal the impact of PA geometry. Let
\begin{IEEEeqnarray}{RCL}
    S_{0,k} &=& \sum_{n=1}^{N}\omega_{k,n},\\
    \mu_{v,k} &=& \frac{1}{S_{0,k}}\sum_{n=1}^{N}\omega_{k,n}v_n,\\
    \sigma_{v,k}^{2} &=& \frac{1}{S_{0,k}}\sum_{n=1}^{N}\omega_{k,n}(v_n-\mu_{v,k})^2.
\end{IEEEeqnarray}
Then, the FIM can be written as
\begin{IEEEeqnarray}{RCL}
    \mathbf J(\mathbf u_k)
    =
    \frac{2S_{0,k}}{\sigma^2}
    \begin{bmatrix}
    u_{x,k}^{2} &
    u_{x,k}(u_{y,k}-\mu_{v,k})\\ \nonumber
    u_{x,k}(u_{y,k}-\mu_{v,k}) &
    (u_{y,k}-\mu_{v,k})^{2}+\sigma_{v,k}^{2}
    \end{bmatrix}.
    \label{eq:FIM_moment_form}
\end{IEEEeqnarray}
This moment-based form shows that localization accuracy improves with the total information weight $S_{0,k}$ and with the effective weighted spatial spread $\sigma_{v,k}^{2}$ of the PA locations. It also indicates that two-dimensional localization becomes ill-conditioned when the user lies close to the waveguide axis, i.e., $u_{x,k}\approx 0$, because the sensitivity in the $x$ direction becomes weak. Therefore, placing PAs with sufficient effective spread along the waveguide and avoiding highly concentrated PA deployments are beneficial for improving localization accuracy.
}
Then, the lower bound of the variance of the estimates $\hat{u}_{x,k}$ and $\hat{u}_{y,k}$ are given as
\begin{IEEEeqnarray}{RCL}
	\V(\hat{u}_{x,k}) & \geq & [{\rm{\mathbf{Cov}}}(\hat{\mathbf{u}}_k)]_{11} = \frac{{J}_{22}}{{J}_{11} {J}_{22} - {J}^2_{12}},\\
	\V(\hat{u}_{y,k}) & \geq & [{\rm{\mathbf{Cov}}}(\hat{\mathbf{u}}_k)]_{22} = \frac{{J}_{11}}{{J}_{11} {J}_{22} - {J}^2_{12}}.
\end{IEEEeqnarray}
Finally, the PEB is expressed as
\begin{equation}
    \mathrm{PEB}_k=\sqrt{\tr\{{\rm{\mathbf{Cov}}}(\hat{\mathbf{u}}_k)\}}=\left(\frac{{J}_{11} + {J}_{22}}{{{J}_{11} {J}_{22} - {J}^2_{12}}}\right)^{\frac{1}{2}},
\end{equation}
where $\tr\{\cdot \}$ denotes the trace operation. 

\section{ML-based User Location Estimation}
As one can see from \eqref{eq:likelihood}, the likelihood function $L_{\mathbf{r}_k}(\mathbf{u}_k)$ is non-convex in the user location $\mathbf{u}_k$, as $\mathbf{u}_k$ appears in non-linear terms in the denominator and the exponent of  $\mathbf{s}_k(\mathbf{{u}_k})$. 
To address it, we employ the following two-step algorithm to find the user location.

First, we discretize  the ground plane into a grid spaced by $d_{\rm{grid}} = \lambda/4$ to obtain initial user locations $\mathbf{u}_{k,\rm{initial}}$. For each initial point $\mathbf{u}_{k,\rm{initial}}$, we calculate the residual error as
\begin{equation*}
\tilde{f}_{\rm error}(\mathbf u_{k,\rm initial})
=
(\mathbf r_k-\mathbf s_k(\mathbf u_{k,\rm initial}))^H
(\mathbf r_k-\mathbf s_k(\mathbf u_{k,\rm initial})).
\end{equation*}
Then, we identify the smallest $N_\mathbf{u}$ values of $\tilde{f}_{\rm{error}}(\mathbf{u}_{k,\rm{initial}})$ such that each one of them will be an input to the second step. {\color{black}This coarse grid search helps identify promising initial points and mitigate the risk of convergence to poor local minima.}

In the second step, we adopt the LM algorithm \cite{levenberg1944method} for each of the $N_\mathbf{u}$ values to iteratively refine the estimates of the user location $\hat{\mathbf{u}}_{k}$. In particular, for iteration $i$, the updated user location is calculated as
\begin{IEEEeqnarray}{RCL}
	\hat{\mathbf{u}}_{k}^{(i+1)} & = & \hat{\mathbf{u}}_{k}^{(i)} + \Delta\hat{\mathbf{u}}_{k}^{(i)},
\end{IEEEeqnarray}
where the update step $\Delta\hat{\mathbf{u}}_{k}^{(i)}$ at iteration $i$ is calculated by solving the following linear system
\begin{IEEEeqnarray}{RCL}
		\left(\frac{\partial \mathbf{s}_k^{\myH}(\hat{\mathbf{u}}_{k}^{(i)})}{\partial \hat{\mathbf{u}}_{k}^{(i)}} \frac{\partial \mathbf{s}_k(\hat{\mathbf{u}}_{k}^{(i)})}{\partial \hat{\mathbf{u}}_{k}^{(i)}}  + \lambda_{\rm{LM}} \mathbf{I} \right)\Delta\hat{\mathbf{u}}_{k}^{(i)} \hspace{3cm} \nonumber \\   \hspace{2cm} =  \Re\left\{\frac{\partial \mathbf{s}_k^{\myH}(\hat{\mathbf{u}}_{k}^{(i)})}{\partial \hat{\mathbf{u}}_{k}^{(i)}} \left[\mathbf{r}_k - \mathbf{s}_k(\hat{\mathbf{u}}_{k}^{(i)})\right]\right\}, \IEEEeqnarraynumspace
\end{IEEEeqnarray}
where $\frac{\partial \mathbf{s}_k(\hat{\mathbf{u}}_{k}^{(i)})}{\partial \hat{\mathbf{u}}}$ is the Jacobian matrix calculated as in \eqref{eq:jac}  and $\lambda_{\rm{LM}}$ is the damping factor.  

{\color{black}
Since the ML objective is non-convex, global optimality over the continuous domain is not guaranteed. Each LM step is accepted only if it decreases the residual objective $\left\|\mathbf r_k-\mathbf s_k(\mathbf u_k)\right\|^2$; hence, accepted iterations monotonically reduce the residual and converge to a stationary point under standard local regularity conditions. The final estimate is selected as the best among the $N_u$ refined candidates.
}




\begin{figure*}[htbp]
\centering
\subfloat[\label{fig:1a}]{\includegraphics[width=0.3\textwidth]{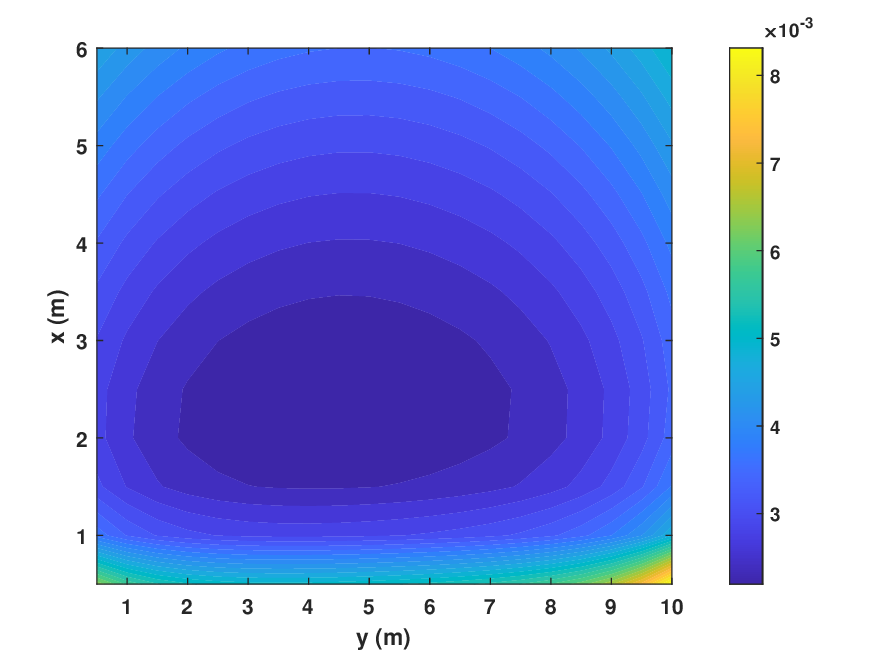}}\hfill
\subfloat[\label{fig:1b}] {\includegraphics[width=0.3\textwidth]{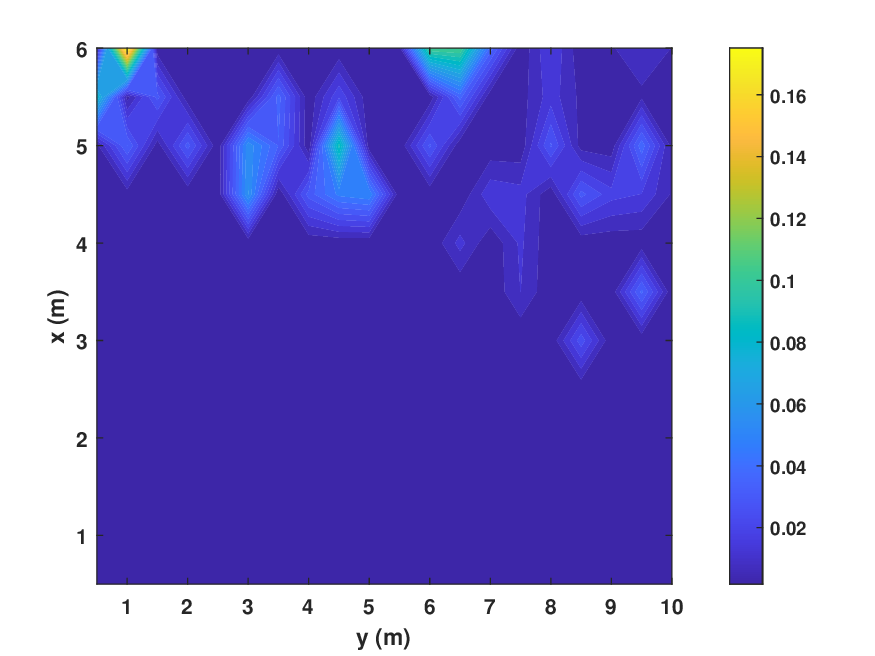}}\hfill
\subfloat[\label{fig:1c}]{\includegraphics[width=0.3\textwidth]{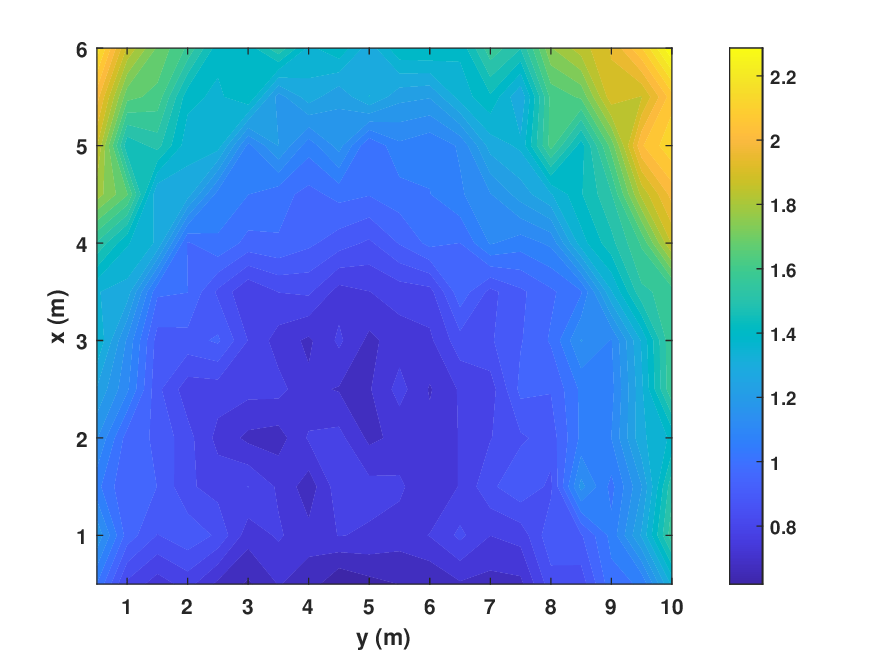}}
 \caption {{\color{black}Position error maps averaged over 100 random trials for: a) PEB; b) proposed solution and c) benchmark from \cite{zhang2025}.}} \label{fig:PEB map}
\end{figure*}

{\color{black}
{\bf{Computational Complexity:}}
Let $N_{\rm grid}$, $N_u$, and $I_{\rm LM}$ denote the number of grid points, retained initial candidates, and maximum LM iterations, respectively. The coarse search requires $\mathcal O(N_{\rm grid}N)$ operations, while selecting the best candidates costs $\mathcal O(N_{\rm grid}\log N_{\rm grid})$. The LM refinement has complexity $\mathcal O(N_u I_{\rm LM}N)$, since each iteration evaluates the signal vector and Jacobian over all PAs and solves a $2\times2$ linear system. Hence, the overall complexity is $\mathcal O(N_{\rm grid}N+N_{\rm grid}\log N_{\rm grid}+N_u I_{\rm LM}N)$.

Note that a smaller grid spacing $d_{\rm grid}$ improves initialization accuracy but increases the complexity of the coarse search. A larger number of retained candidates $N_u$ improves robustness against poor initialization but increases the LM refinement cost. The damping factor $\lambda_{\rm LM}$ controls the transition between Gauss--Newton-like steps and more conservative gradient-descent-like steps. The maximum number of LM iterations $I_{\rm LM}$ controls the refinement accuracy and runtime.  
}





\section{Numerical Results}
Numerical results are presented in this section to evaluate the performance of the proposed phase-aware user position estimation method. 
Unless otherwise specified, the simulation parameters are as follows \cite{zhang2025}: the considered deployment area is a rectangular region of size $6 \times 10$ m$^2$, and the waveguide height is set to $d = 3$ m. The subcarrier frequency is $2.8$ GHz, and the user transmit power is fixed at $p_k = 0.1$ W. {\color{black}The number of PAs is $N=8$, and the noise power is -50 dBm.} The relative permittivity and dielectric loss tangent are $\epsilon_r = 2.08$ and $\tan(\delta) = 0.0004$, respectively, following \cite{rao2015microwave}. The number of initially selected user locations is $N_\mathbf{u}=20$. 

For comparison, we consider {\color{black}four} benchmarks: the amplitude-only weighted least squares (WLS) method in \cite{zhang2025}, an RSS-only grid-search estimator, a phase-only estimator, and {\color{black}a conventional fixed-antenna ULA baseline. The fixed-antenna baseline uses the same phase-aware ML estimator as the proposed method, but the antennas are fixed at the feed point with half-wavelength spacing.}

Fig. \ref{fig:performance} compares the average position error under different system parameters. In Fig. \ref{fig:performance}(a), the error increases with the noise power for all schemes, since the amplitude and phase observations become less reliable. The proposed ML estimator achieves the lowest error, consistent with \eqref{eq:omega_decomposition}--\eqref{ratio}, where the phase-induced information provides a dominant contribution to the FIM. Fig. \ref{fig:performance}(b) shows that increasing the number of PAs improves localization accuracy by increasing the total Fisher information and the effective PA spatial diversity. Fig. \ref{fig:performance}(c) evaluates the effect of user--PA distance. The user locations are first randomly generated, and their $x$-coordinates are then shifted by the distance shown on the x-axis. The error increases with this shift because the received signal and geometric sensitivity become weaker. In all cases, the proposed method outperforms the amplitude-only WLS/RSS benchmarks and the phase-only estimator, confirming the benefit of jointly exploiting amplitude and phase. {\color{black}Compared with the fixed-antenna baseline, PASS benefits from a larger effective aperture and improved geometry-dependent spatial diversity, while the fixed array avoids in-waveguide attenuation; hence, the results also illustrate the practical trade-off between spatial flexibility and waveguide propagation loss.}

{\color{black}Fig.~\ref{fig:waveguid loss} evaluates the impact of the dielectric loss tangent $\tan(\delta)$ on the localization accuracy. As $\tan(\delta)$ increases, the in-waveguide attenuation becomes stronger, which reduces the effective contribution of PAs farther from the feed point. Consequently, the positioning error increases for all PASS-based methods. Nevertheless, the proposed estimator maintains the lowest error, showing its robustness to waveguide attenuation within the considered range.}

Fig. \ref{fig:PEB map} shows heat maps of the positioning performance over the considered area, where darker colors indicate lower error and brighter colors indicate larger error. Fig. \ref{fig:PEB map}(a) plots the PEB obtained from the derived CRLB. As predicted by the moment-based FIM, the PEB increases when $u_{x,k}$ approaches zero, since the sensitivity in the $x$-direction becomes weak and the localization problem becomes ill-conditioned. Fig. \ref{fig:PEB map}(b) presents the empirical error of the proposed estimator, which remains relatively small over most of the considered area. In contrast, the amplitude-only benchmark in Fig. \ref{fig:PEB map}(c) suffers from much larger errors, especially in unfavorable regions. These results verify that exploiting the complex signal phase substantially improves localization accuracy compared with amplitude-only localization.

\section{Conclusion}
{\color{black}This letter studied phase-aware user localization in PASS by exploiting the full complex baseband signal. A unified signal model was developed, and closed-form FIM, CRLB, and PEB expressions were derived, revealing that the phase-induced information decays more slowly with distance than the amplitude-induced information. A two-stage ML estimator based on coarse grid search and LM refinement was then proposed. 
Numerical results confirmed that the proposed estimator achieves low positioning error over most of the considered area and significantly outperforms amplitude-only benchmarks under different noise levels, PA numbers, and user locations.}
    
\bibliographystyle{IEEEtran}
\bibliography{biblio}

\balance
\end{document}